\documentclass[showpacs,preprintnumbers,amsmath,amssymb,prb]{revtex4}

\usepackage{graphicx}
\usepackage{dcolumn}
\usepackage{bm}

\begin{document}

\title{Dynamics of spin-$\frac12$ $J_1$--$J_2$ model on the triangular lattice}

\author{A.\ V.\ Syromyatnikov}
\email{asyromyatnikov@yandex.ru}
\affiliation{Petersburg Nuclear Physics Institute named by B.P.Konstantinov of National Research Center "Kurchatov Institute", Gatchina 188300, Russia}

\date{\today}

\begin{abstract}

We discuss spin-$\frac12$ $J_1$--$J_2$ model on the triangular lattice using recently proposed bond-operator theory (BOT). In agreement with previous discussions of this system, we obtain four phases upon $J_2$ increasing: the phase with $120^\circ$ ordering of three sublattices, the spin-liquid phase, the state with the collinear stripe order, and the spiral phase. The $120^\circ$ and the stripe phases are discussed in detail. All calculated static characteristics of the model are in good agreement with previous numerical findings.
In the $120^\circ$ phase, we observe the evolution of quasiparticles spectra and dynamical structure factors (DSFs) upon approaching the spin-liquid phase. Some of the considered elementary excitations were introduced first in our recent study of this system at $J_2=0$ using the BOT. 
In the stripe phase, we observe that the doubly degenerate magnon spectrum known from the spin-wave theory (SWT) is split by quantum fluctuations which are taken into account more accurately in the BOT. As compared with other known findings of the SWT in the stripe state, we observe additional spin-1 and spin-0 quasiparticles which give visible anomalies in the transverse and longitudinal DSFs. We obtain also a special spin-0 quasiparticle named singlon who produces a peak only in four-spin correlator and who is invisible in the longitudinal DSF. We show that the singlon spectrum lies below energies of all spin-0 and spin-1 excitations in some parts of the Brillouin zone. Singlon spectrum at zero momentum can be probed by the Raman scattering.

\end{abstract}


\maketitle

\section{Introduction}

\begin{figure}
\includegraphics[scale=0.75]{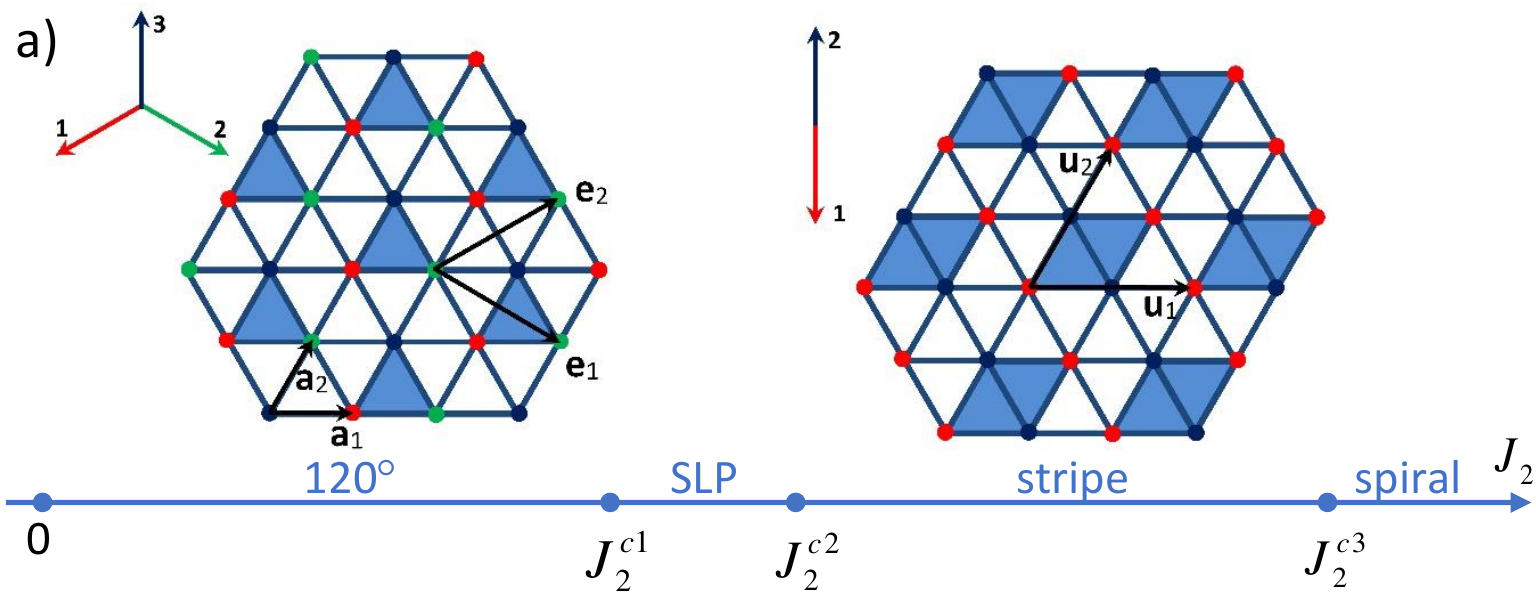}
\includegraphics[scale=0.75]{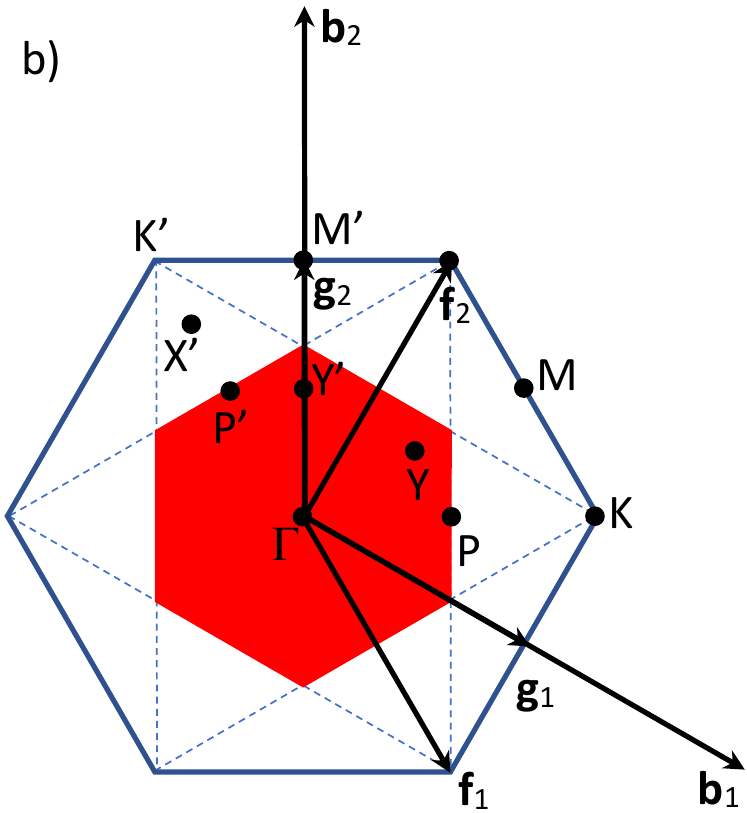}
\caption{
(a) Phases of $J_1$--$J_2$ model \eqref{ham} on the triangular lattice: three-sublattice state with $120^\circ$ magnetic ordering ($120^\circ$), spin-liquid phase (SLP), two-sublattice collinear stripe phase (stripe), and the state with an incommensurate spiral ordering (spiral). Sites are distinguished by color belonging to different magnetic sublattices in the $120^\circ$ and in the stripe phases. Translation vectors are shown of the crystal (${\bf a}_{1,2}$), of the magnetic lattice in the $120^\circ$ state (${\bf e}_{1,2}$), and of the magnetic lattice in the stripe phase with the extended unit cell considered within the BOT (${\bf u}_{1,2}$). It was found before that $J_2^{c1}\approx0.07$, $J_2^{c2}\approx0.15$, and $J_2^{c3}\approx1$. \cite{trij1j2,triang3,triang4,oitmaatri,CCMtri,gsj1j21,gsj1j22,sl1,sl2}
(b) Translation vectors ${\bf b}_{1,2}$, ${\bf f}_{1,2}$, and ${\bf g}_{1,2}$ are depicted of reciprocal lattices corresponding to ${\bf a}_{1,2}$, ${\bf e}_{1,2}$, and ${\bf u}_{1,2}$, respectively. Blue hexagon is the crystal Brillouin zone. Red hexagon is the magnetic Brillouin zone for the $120^\circ$ state. Some high-symmetry points in the crystal Brillouin zone are shown. Dashed lines indicate high-symmetry directions which are important for the present consideration.
\label{lattfig}}
\end{figure}

Spin-$\frac12$ Heisenberg antiferromagnet on the triangular lattice has attracted much attention since the paper by P.\ W.\ Anderson \cite{anderson1} highlighting the role of frustration in stabilization of quantum spin-liquid phases (SLPs) in systems on lattices with spatial dimensions greater than one. The most subsequent analytical and numerical works show that the $120^\circ$ magnetically ordered state with three sublattices is stable in this model. \cite{egs1,egs2,egs3} However, this system is not far from a SLP which can be stabilized by introducing to the Hamiltonian
\begin{equation}
\label{ham}
{\cal H} = \sum_{\langle i,j \rangle}	{\bf S}_i{\bf S}_j + J_2\sum_{\langle \langle i,j \rangle\rangle}	{\bf S}_i{\bf S}_j
\end{equation} 
a small next-nearest-neighbor exchange coupling $J_2>0$, where the first term describes the nearest-neighbor spin interaction in which we put $J_1=1$. It was found \cite{trij1j2,triang3,triang4,oitmaatri,CCMtri,gsj1j21,gsj1j22,sl1,sl2} that the $J_1$--$J_2$ model shows four phases presented schematically in Fig.~\ref{lattfig}(a), where the SLP arising at $0.07\alt J_2\alt0.15$ is sandwiched between phases with $120^\circ$ and stripe collinear magnetic orders and an incommensurate spiral ordering is stabilized at $J_2>1$. The nature of the SLP is debated now: numerical evidences are reported for gapped \cite{sl1,sl2} and gapless \cite{triang3,triang4} spin liquids.

Dynamical properties of the $120^\circ$ ordered state has attracted great interest recently because it turned out that standard analytical approaches failed to describe even qualitatively the short-wavelength spin dynamics in model \eqref{ham} at $J_2=0$. In particular, neutron scattering experiments \cite{triang1,bacoprl,bacoprb} carried out in $\rm Ba_3CoSb_2O_9$, which is described well by Eq.~\eqref{ham} with small easy-plane anisotropy and $J_2=0$, show at least four peaks at $M$ point (see Fig.~\ref{lattfig}(b)) of the Brillouin zone (BZ). In contrast, the spin-wave theory (SWT) predicts only two magnon peaks at $M$ and a high-energy continuum of excitations. \cite{chub_triang,zh_triang,zhito} These experimentally observed anomalies are reproduced quantitatively numerically using the tensor network renormalization group method. \cite{navy} Recent application of the Schwinger boson approach to this problem reproduces qualitatively high-energy peculiarities in experimental data. \cite{batista} 

We have attacked recently the problem of the short-wavelength spin dynamics in model \eqref{ham} at $J_2=0$ using the bond-operator technique (BOT) proposed in Ref.~\cite{ibot} which is discussed briefly in Sec.~\ref{bot}. \cite{itri} We have obtained that quantum fluctuations considerably change properties of three conventional magnon modes predicted by the SWT. \cite{itri} In particular, we have found that in agreement with the experiment in $\rm Ba_3CoSb_2O_9$ quantum fluctuations lift the degeneracy between two magnon modes at $M$ point predicted by the SWT. Besides, we have observed novel high-energy collective excitations built from high-energy excitations of the magnetic unit cell (containing three spins) and another novel high-energy quasiparticle which has no counterpart not only in the SWT but also in the harmonic approximation of the BOT. All observed elementary excitations produce visible anomalies in dynamical spin correlators and describe experimental data obtained in $\rm Ba_3CoSb_2O_9$. \cite{itri}

In the present study, we continue our consideration of triangular-lattice antiferromagnets by the BOT and trace the spectra evolution in the $J_1$--$J_2$ model upon variation of $J_2>0$ in the $120^\circ$ and stripe ordered phases. We discuss static properties in Sec.~\ref{stat} and show that the staggered magnetization obtained using the BOT follows quite accurately previous numerical findings with the interval of the non-magnetic phase stability being $0.1<J_2<0.16$. We discuss dynamical properties of model \eqref{ham} in Sec.~\ref{dyn}. In the $120^\circ$ phase, we demonstrate that the continuum of excitations moves closer to the lowest magnon mode upon $J_2$ increasing in agreement with previous numerical findings \cite{triang3}. The remaining two conventional magnon modes acquire noticeable damping on the way to the SLP while some other modes found in Ref.~\cite{itri} remain well-defined and produce visible high-energy anomalies in the dynamical structure factor (DSF).

In the collinear stripe phase which has been much less studied before, we find that quantum fluctuations split the magnon spectrum that is doubly degenerate according to the semi-classical SWT. The value of this splitting is noticeable for short-wavelength magnons. We find also low-lying well-defined spin-0 excitations producing visible anomalies in the longitudinal DSF which can be observed experimentally. Spectra of these spin-0 excitations become closer to magnon spectra on the way to the SLP. Besides, we find a special spin-0 elementary excitation which is seen in the harmonic approximation of the BOT as a singlet spin state of the magnetic unit cell propagating along the lattice. We call this quasiparticle "singlon" in Ref.~\cite{ibot}. Such excitation is invisible for neutrons. It appears also in square-lattice spin-$\frac12$ Heisenberg antiferromagnet and produces a broad anomaly (known in the literature as "two-magnon peak") in the Raman spectra in the $B_{1g}$ geometry which was observed, e.g., in layered cuprates.

Sec.~\ref{conc} contains a summary and our conclusion.

\section{Bond operator technique for $J_1$--$J_2$ model on the triangular lattice}
\label{bot}

\begin{figure}
\includegraphics[scale=0.4]{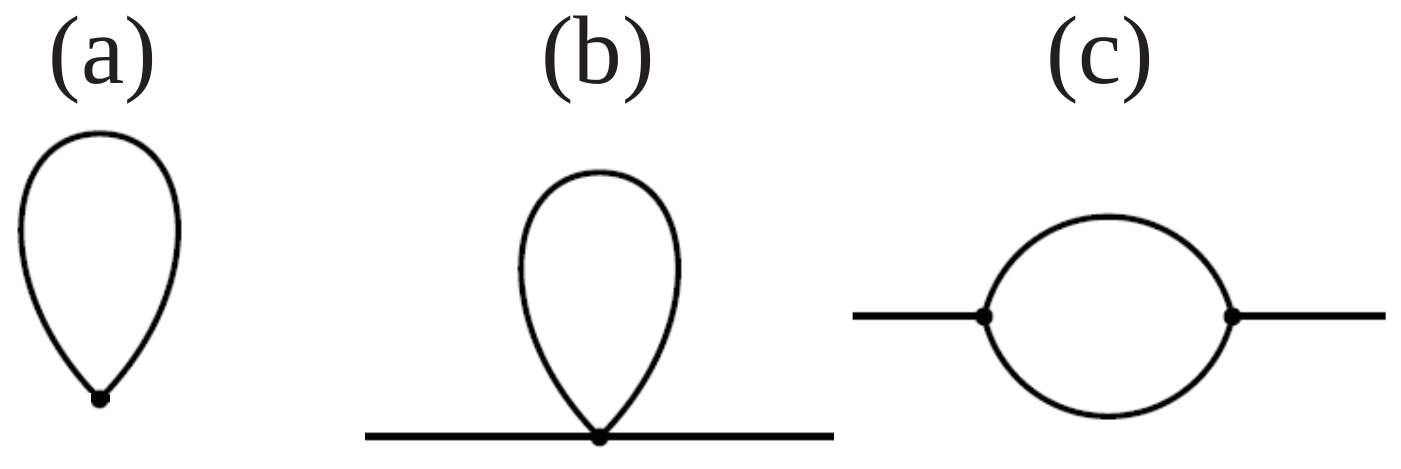}
\caption{Diagrams giving corrections of the first-order in $1/n$ to (a) the ground state energy and the staggered magnetization, and (b), (c) to self-energy parts.
\label{diag}}
\end{figure}

We use in the present study the bond-operator technique proposed and discussed in detail in Ref.~\cite{ibot}. The main idea of this approach is to take into account all spin degrees of freedom in the magnetic unit cell containing several spins 1/2 by building a bosonic spin representation reproducing the spin commutation algebra. A general scheme of construction of such representation for arbitrary number of spins in the unit cell is described in detail in Ref.~\cite{ibot}. We consider now briefly the main steps of this procedure by the example of three spins in the unit cell which is relevant for the $120^\circ$ phase (see Fig.~\ref{lattfig}). First, we introduce seven Bose operators in each unit cell which act on eight basis functions of three spins $|0\rangle$ and $|e_i\rangle$ ($i=1,...,7$) according to the rule
\begin{equation}
\label{bosons}
	a_i^\dagger |0\rangle = |e_i\rangle, \quad i=1,...,7,
\end{equation}
where $|0\rangle$ is a selected state playing the role of the vacuum. Then, we build the bosonic representation of spins in the unit cell as it is described in Ref.~\cite{ibot} which turns out to be quite bulky so that we do not present it here. The code in the Mathematica software which generates this representation is presented in the Supplemental Material \cite{supp1}. There is a formal artificial parameter $n$ in this representation that appears in operator $\sqrt{n-\sum_{i=1}^7a_i^\dagger a_i}$ by which linear in Bose operators terms are multiplied (cf.\ the term $\sqrt{2S-a_i^\dagger a_i}$ in the Holstein-Primakoff representation). It prevents mixing of states containing more than $n$ bosons and states with no more than $n$ bosons (then, the physical results of the BOT correspond to $n=1$). Besides, all constant terms in our representation of spin components are proportional to $n$ whereas bilinear in Bose operators terms do not depend on $n$ and have the form $a_i^\dagger a_j$. We introduce also separate representations via operators \eqref{bosons} for terms ${\bf S}_i{\bf S}_j$ in the Hamiltonian in which $i$ and $j$ belong to the same unit cell. Constant terms in these representations are proportional to $n^2$ and terms of the form $a_i^\dagger a_j$ are proportional to $n$. \cite{ibot} Thus, we obtain a close analog of the conventional Holstein-Primakoff spin transformation which reproduces the commutation algebra of all spin operators in the unit cell for all $n>0$ and in which $n$ is the counterpart of the spin value $S$. In analogy with the SWT, expressions for observables are found in the BOT using the conventional diagrammatic technique as series in $1/n$. This is because terms in the Bose-analog of the spin Hamiltonian containing products of $i$ Bose operators are proportional to $n^{2-i/2}$ (in the SWT, such terms are proportional to $S^{2-i/2}$). For instance, to find the ground-state energy, the staggered magnetization, and self-energy parts in the first order in $1/n$ one has to calculate diagrams shown in Fig.~\ref{diag} (as in the SWT in the first order in $1/S$). 

Our previous applications of the BOT to two-dimensional spin-$\frac12$ models well studied before by other numerical and analytical methods show that first $1/n$ terms in most cases give the main corrections to renormalization of observables if the system is not very close to a quantum critical point (similarly, first $1/S$ corrections in the SWT frequently make the main quantum renormalization of observable quantities even at $S=1/2$, Ref.~\cite{monous}). \cite{ibot,aktersky} Importantly, because the spin commutation algebra is reproduced in our method at any $n>0$, the proper number of Goldstone excitations arises in ordered phases in any order in $1/n$ (unlike the vast majority of other versions of the BOT proposed so far \cite{ibot}). Although the BOT is technically very similar to the SWT, the main disadvantage of this technique is that it is very bulky (e.g., the part of the Hamiltonian bilinear in Bose operators contains more than 100 terms) and it requires time-consuming numerical calculation of diagrams. That is why there is a limited number of points on some plots below found in the first order in $1/n$.

The construction of the four-spin variant of the BOT is discussed in detail in Ref.~\cite{ibot}, where the spin representation is presented explicitly. The code in the Mathematica software which generates this representation together with the bilinear part of the Hamiltonian is presented in the Supplemental Material \cite{supp2}. We use below the three-spin and the four-spin variants of the BOT for consideration of the $120^\circ$ and the stripe phases, respectively. Then, we use the unit cell in the consideration of the stripe phase which is two times as large as the magnetic unit cell. The extension of the unit cell is very useful in the BOT because it allows to consider numerous interesting excitations which can arise in standard approaches as bound states of conventional quasiparticles (magnons or triplons). \cite{ibot} Consideration of the bound states require analysis of some infinite series of diagrams in common methods. In contrast, there are separate bosons in the BOT describing some of them that allows, in particular, to find their spectra as series in $1/n$ by calculating the same diagrams as for the common quasiparticles (e.g., diagrams shown in Figs.~\ref{diag}(b) and \ref{diag}(c) in the first order in $1/n$). As it is discussed in more detail in Ref.~\cite{ibot}, the version of the BOT with two-site unit cell contains three bosons describing in the ordered phase two spin-1 excitations (conventional magnons) and one spin-0 quasiparticle (the Higgs mode). In contrast, the four-site version of the BOT contains 15 bosons describing (along with conventional magnons and the Higgs mode), in particular, the "singlon" whose energy is lower than energies of all quasiparticles in some parts of the BZ (see below) and who can be probed by the Raman scattering. The price to pay for the increasing of the quasiparticles zoo is the bulky theory.

We take into account below diagrams shown in Fig.~\ref{diag}(b) and \ref{diag}(c) to find all self-energy parts $\Sigma(\omega,{\bf k})$ in the first order in $1/n$. We use (bare) Green's functions of the harmonic approximation in these calculations. Spectra of elementary excitations are obtained in two ways below. First, by expanding Green's functions denominators near a bare spectrum up to the first order in $1/n$ and putting $\omega$ equal to the bare spectrum in self-energy parts. This is a usual way of finding spectra in the first order in the expansion parameter ($1/n$ in this case). In particular, spectra of all Goldstone quasiparticles calculated in this way remain gapless as it is noted above. Second, we find zeros of Green's functions denominators by taking into account $\omega$-dependence of self-energy parts (a self-consistent scheme). Results obtained in these two schemes are different. The difference is usually small for low-energy quasiparticles whereas the difference can be large for the high-energy short-wavelength elementary excitations. \cite{ibot,iboth,itri} The self-consistent scheme is usually applied in various theoretical considerations when first-order corrections renormalize bare spectra considerably (see, e.g., Ref.~\cite{zhito}). The self-consistent scheme can even change qualitatively the physical picture by revealing new poles of the Green's functions. Such results should be treated with caution and should be corroborated by numerical and/or experimental data. However our previous applications of the BOT to other systems show that the self-consistent scheme give results in the high-energy sector which are in a good agreement with numerical and experimental findings. \cite{ibot,iboth,itri} In particular, in agreement with previous numerical findings, we demonstrated using the BOT that instead of one magnon peak predicted by the linear SWT numerous high-energy anomalies arise in dynamical structure factors (DSFs) in Heisenberg antiferromagnet in strong magnetic field below its saturation value. \cite{iboth} 

DSFs presented below are also obtained by taking into account the $\omega$-dependence of self-energy parts (as in Refs.~\cite{iboth,itri}). We discuss below anomalies in DSFs corresponding to zeros of Green's functions denominators which are found using the self-consistent scheme and indicated in insets of corresponding plots.

\section{Static properties}
\label{stat}

\begin{figure}
\includegraphics[scale=0.9]{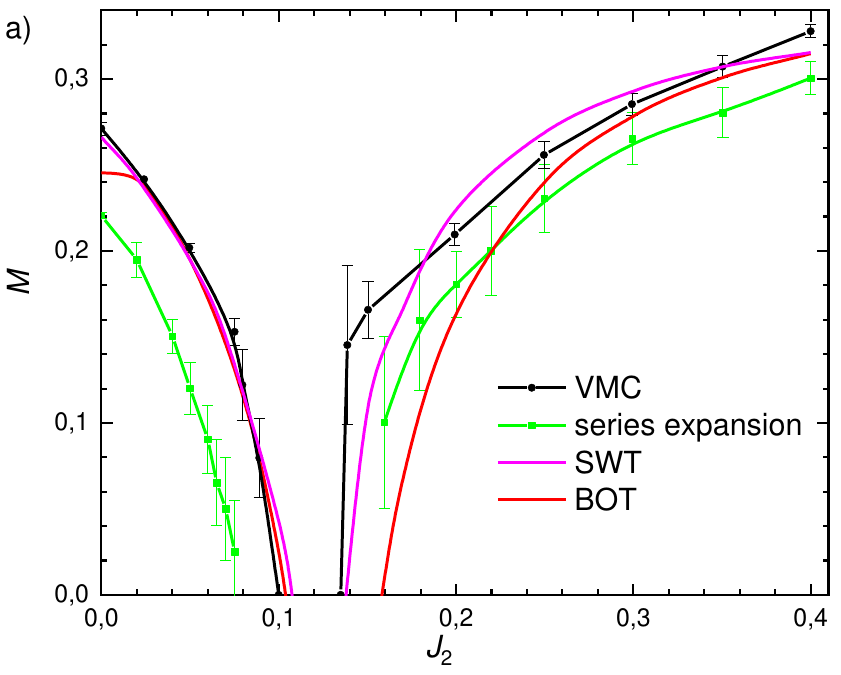}
\includegraphics[scale=0.9]{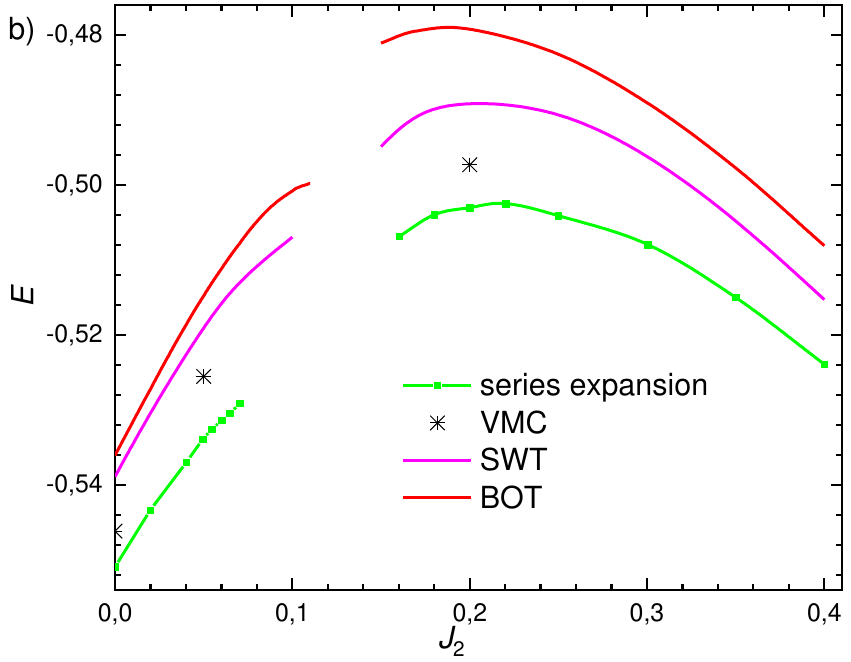}
\caption{(a) Staggered magnetization per site $M$ and (b) the ground-state energy per spin $E$ obtained in the first order in $1/S$ (SWT), using the variational Monte Carlo (VMC) \cite{trij1j2,triang4}, the series expansion \cite{oitmaatri}, and the bond-operator technique (BOT) in the first order in $1/n$ (present study). Results for $E$ found using the coupled cluster method \cite{CCMtri} are indistinguishable from the series expansion results. 
\label{ME}}
\end{figure}

Graphics are shown in Fig.~\ref{ME} of the staggered magnetization per site $M$ and the ground-state energy per spin $E$ obtained in the first order in $1/n$ as it is explained in detail in Refs.~\cite{ibot,itri}. It is seen that our results for $M$ are very close to previous numerical findings and give for the region of the SLP stability 
\begin{equation}
\label{slreg}
	0.1<J_2<0.16
\end{equation}
in agreement with many previous results \cite{sl1,sl2,triang3,triang4}. It is also seen from Fig.~\ref{ME}(b) that the BOT overestimates the ground-state energy by 2-6 \% in the considered model.

\section{Dynamical properties}
\label{dyn}

\subsection{$120^\circ$ phase} 

We calculate in this section the dynamical spin susceptibility
\begin{equation}
\label{chi}
\chi({\bf k},\omega) = 
i\int_0^\infty dt 
e^{i\omega t}	
\left\langle \left[ {\bf S}_{\bf k}(t), {\bf S}_{-\bf k}(0) \right] \right\rangle
\end{equation}
and the dynamical structure factor (DSF)
\begin{equation}
\label{dsf}
{\cal S}({\bf k},\omega) = 
\frac1\pi {\rm Im}
\chi({\bf k},\omega).
\end{equation}
In the $120^\circ$ phase, ${\bf S}_{\bf k}$ are built on spin operators 1, 2, and 3 in the magnetic unit cell (see Fig.~\ref{lattfig}(a)) as follows:
\begin{equation}
\label{sk}
	{\bf S}_{\bf k} = \frac{1}{\sqrt3} 
	\left( 
	{\bf S}_{1\bf k} + {\bf S}_{2\bf k}e^{-i(k_1+k_2)/3} + {\bf S}_{3\bf k}e^{-i(2k_2-k_1)/3}
	\right),
\end{equation}
where ${\bf k} = k_1{\bf f}_1 + k_2{\bf f}_2$, and ${\bf f}_{1,2}$ are depicted in Fig.~\ref{lattfig}(b). 

\begin{figure}
\includegraphics[scale=0.9]{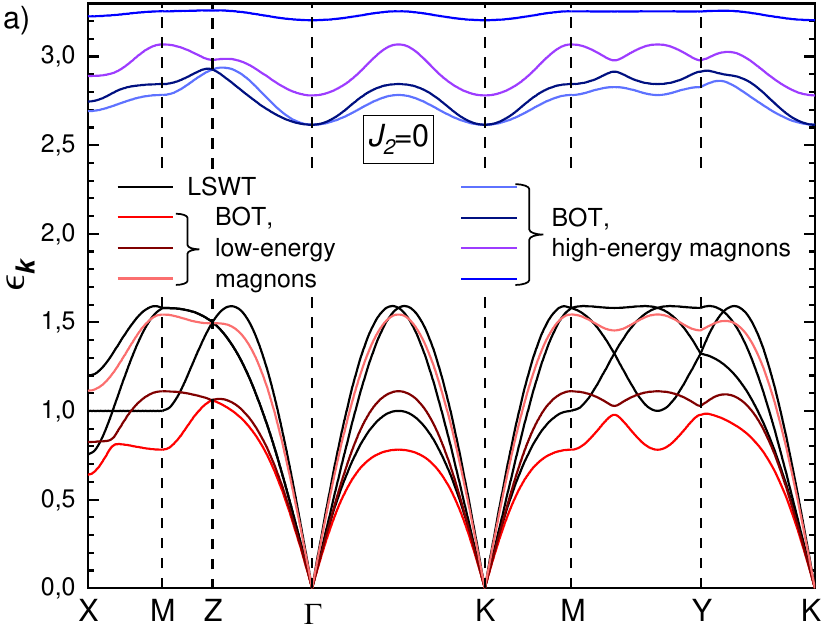}
\includegraphics[scale=0.9]{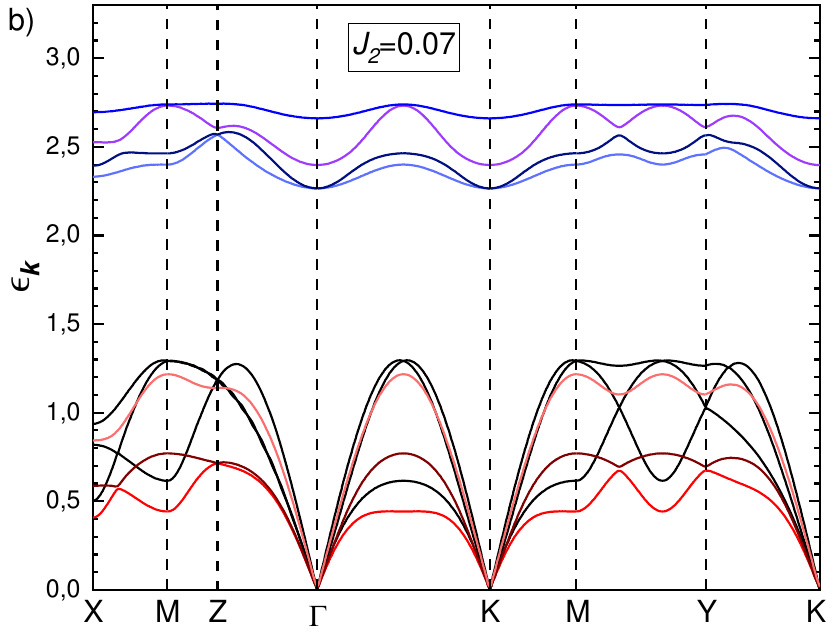}
\caption{
Spectra of elementary excitations corresponding to poles of dynamical spin susceptibility \eqref{chi} obtained in the $120^\circ$ phase at (a) $J_2=0$ and (b) $J_2=0.07$ in the linear spin-wave theory (LSWT) and within the harmonic approximation of the bond-operator technique (BOT). The contour passes through high-symmetry points of the Brillouin zone shown in Fig.~\ref{lattfig}(b).
\label{spec0}}
\end{figure}

Spectra of all elementary excitations are shown in Fig.~\ref{spec0} obtained in the harmonic approximation of the BOT at $J_2=0$ and $J_2=0.07$. Spectra at $J_2=0$ are discussed in detail in Ref.~\cite{itri}. All elementary excitations produce anomalies in DSF \eqref{dsf}. Then, we call three Goldstone excitations (which are known, e.g., from the SWT) "low-energy magnons" and the rest four "optical" branches are named "high-energy magnons". By comparing results in Fig.~\ref{spec0} obtained in the BOT and in the linear SWT, one notes that quantum fluctuations strongly modify and move down spectra of low-energy magnons. The most striking difference between predictions of these approaches is that quantum fluctuations (which are taken into account in the BOT more accurately) remove the degeneracy between two magnon branches predicted by the semi-classical SWT along $\Gamma M$ and dashed lines shown in Fig.~\ref{lattfig}(b). We show in Ref.~\cite{itri} that this our finding is in a quantitative agreement with experimental data in $\rm Ba_3CoSb_2O_9$. It is also seen from Fig.~\ref{spec0} that all spectra in the BOT move down upon $J_2$ increasing.

\begin{figure}
\includegraphics[scale=0.9]{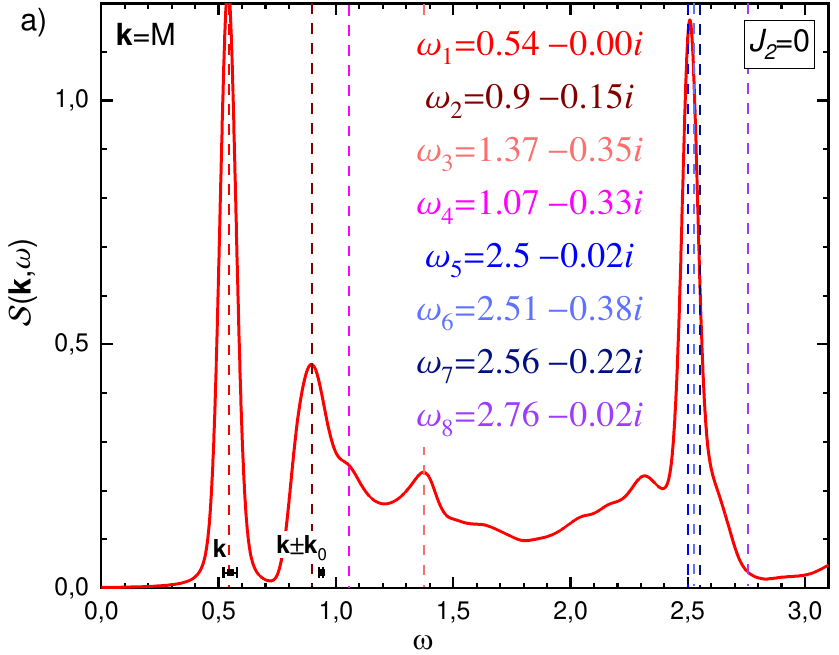}
\includegraphics[scale=0.9]{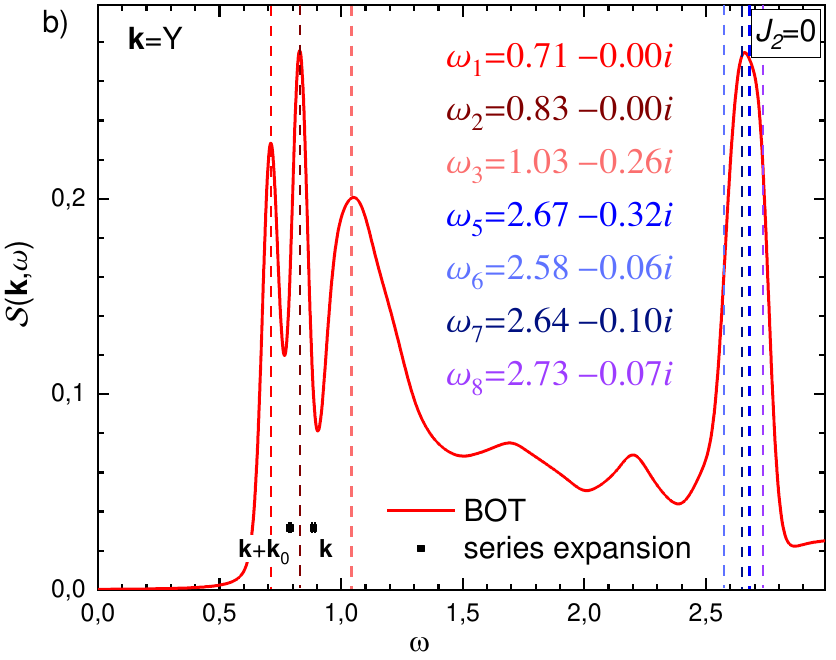}
\includegraphics[scale=0.9]{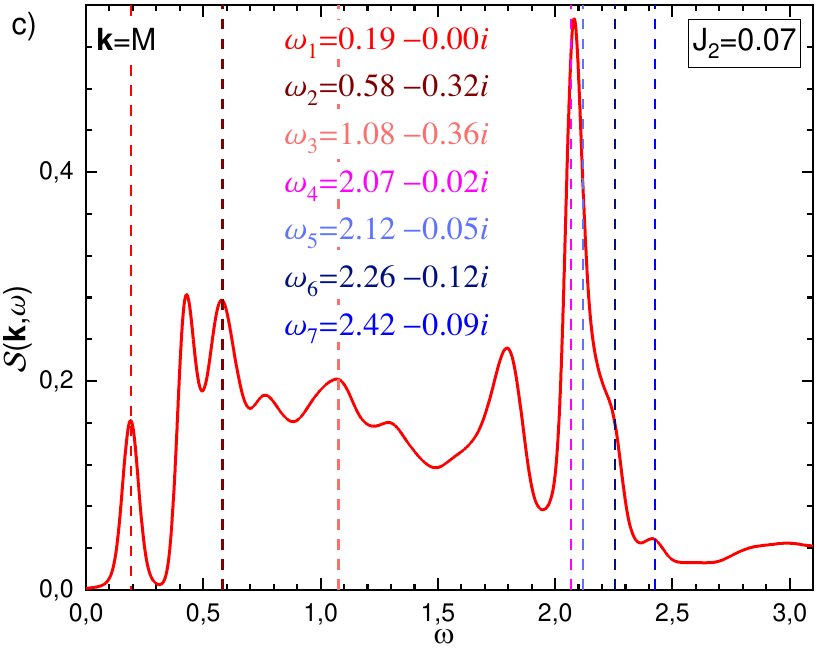}
\includegraphics[scale=0.9]{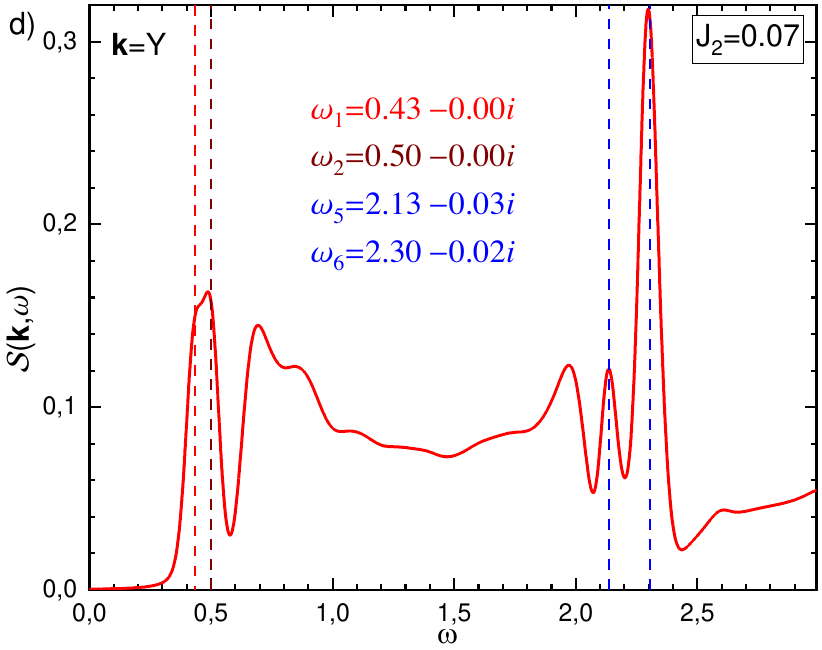}
\caption{
Dynamical structure factor (DSF) at points $M$ and $Y$ of the BZ (see Fig.~\ref{lattfig}(b)). DSF obtained within the first order in $1/n$ has been convoluted with the energy resolution of $0.03$. Magnon energies are also indicated in panels (a) and (b) which were obtained in Ref.~\cite{series_tri} using the series expansion technique. Anomalies in the DSF are produced by poles of spin correlator \eqref{chi} indicated in insets by colors corresponding to excitation branches shown in Fig.~\ref{spec0}. Real parts of these poles are marked by vertical dashed lines of respective colors. Imaginary parts of poles correspond to quasiparticles damping. Poles are not shown whose imaginary parts exceed one third of their real parts. Pole $\omega_4$ in panels (a) and (c) has no counterpart neither in the spin-wave theory nor in the harmonic approximation of the BOT.
\label{dsfmy}}
\end{figure}

Corrections to self-energy parts of the first order in $1/n$ renormalize quasiparticles energies, lead to a finite damping of some of them, give rise an incoherent background in DSFs, and can produce novel poles in Green's functions which have no counterparts neither in the SWT nor in the harmonic approximation of the BOT. An anomaly produced by such novel pole is clearly seen in the DSF at $M$ point near $\omega\approx1.1$ (see Fig.~\ref{dsfmy}(a)). Although the imaginary part of this pole is quite large at $J_2=0$, it can be reduced to zero by introducing to the model a small easy-plane anisotropy. \cite{itri} As a result, the novel quasiparticle corresponding to this pole becomes well-defined and produces an anomaly in the DSF which, as we propose in Ref.~\cite{itri}, is observed in neutron experiments in $\rm Ba_3CoSb_2O_9$. It is also seen in Figs.~\ref{dsfmy}(a) and \ref{dsfmy}(b) that high-energy magnons produce high-energy anomalies in the DSF which is also observed experimentally in $\rm Ba_3CoSb_2O_9$ (see Ref.~\cite{itri}). 

Figs.~\ref{dsfmy}(c) and \ref{dsfmy}(d) illustrate the evolution of the DSF at $M$ and $Y$ points upon $J_2$ increasing (cf. Figs.~\ref{dsfmy}(a), \ref{dsfmy}(b) and \ref{dsfmy}(c), \ref{dsfmy}(d)). All magnon poles move to lower energies and imaginary parts of almost all of them increase. Some magnons become badly defined and anomalies from them are washed out (we do not indicate poles in Fig.~\ref{dsfmy} whose imaginary parts exceed one third of their real parts). The high-energy anomaly produced by high-energy magnons remains as $J_2$ rises. The damping of the novel quasiparticle increases quickly at $M$ as $J_2$ rises from zero. Then, it is difficult to trace its evolution. However, we can identify a novel pole of the dynamical spin susceptibility \eqref{chi} at $\omega\approx2.07$ and $J_2=0.07$ which is shown in Fig.~\ref{dsfmy}(c). The lower edge of the incoherent background moves down upon $J_2$ increasing and it is marked by a small anomaly in DSFs (at $\omega\approx0.45$ and $\omega\approx0.7$ for $M$ and $Y$ points, respectively). It is shown numerically in Ref.~\cite{triang3} that this continuum of excitations merges with the lower magnon branch at the quantum critical point (QCP), where it is better represented as a two-spinon continuum. In agreement with both SWT \cite{gsj1j21,gsj1j22} and previous numerical findings \cite{triang3}, we obtain that magnon spectrum becomes soft at the $M$ point at the QCP.

Our results shown in Figs.~\ref{dsfmy}(c) and \ref{dsfmy}(d) are in overall agreement with neutron data observed in $\rm KYbSe_2$ that is believed to be described by model \eqref{ham} with $J_2\approx0.05$ and $J_1\approx0.56$~meV. \cite{kybse} In particular, a broad continuum is seen in experimental data at $M$ point which starts at $\approx$0.2~meV$\approx0.36J_1$ with a small peak and extends up to $\approx$1.4~meV$\approx2.5J_1$ (see Figs.~5 and S4 in Ref.~\cite{kybse}). There is also a broad anomaly at $\approx$0.8~meV$\approx1.4J_1$ at $M$ which probably stems from the high-energy broad peak observed experimentally \cite{triang1,bacoprl,bacoprb} and numerically \cite{navy} in $\rm Ba_3CoSb_2O_9$ at $\omega\approx1.9$ and $J_2=0$. Our calculations show that the weak low-energy anomaly arises at $M$ at $\omega\approx0.36$ when $J_2\approx0.035$. We propose that the broad anomaly seen in $\rm KYbSe_2$ at $\approx$0.8~meV is produced by the high-energy magnons which produce also the high-energy broad peak in $\rm Ba_3CoSb_2O_9$ (see Ref.~\cite{itri} for extra detail). The position of this anomaly is overestimated in the first order of the BOT by $\approx30\%$ both in $\rm Ba_3CoSb_2O_9$ and $\rm KYbSe_2$. A less intense continuum is also seen in $\rm KYbSe_2$ at $Y$ point within approximately the same energy interval as at $M$ that is in an overall agreement with Fig.~\ref{dsfmy}(d). Our results are also in a qualitative agreement with those obtained by the Schwinger boson approach in Ref.~\cite{kybse}.


\subsection{Stripe phase}

As soon as the longitudinal and the transverse channels are separated in the collinear stripe phase, it is reasonable to introduce the following dynamical spin susceptibilities:
\begin{eqnarray}
\label{chistr}
\chi_{\alpha\beta}({\bf k},\omega) &=&
i\int_0^\infty dt 
e^{i\omega t}	
\left\langle \left[ S^\alpha_{\bf k}(t), S^\beta_{-\bf k}(0) \right] \right\rangle,\\
\label{chizz}
\chi_\|({\bf k},\omega) &=& \chi_{zz}({\bf k},\omega),\\
\label{chiperp}
\chi_\perp({\bf k},\omega) &=& \chi_{xx}({\bf k},\omega) +  \chi_{yy}({\bf k},\omega),
\end{eqnarray}
and consider longitudinal and transverse DSFs
\begin{eqnarray}
\label{dsfzz}
{\cal S}_\|({\bf k},\omega) &=& 
\frac1\pi {\rm Im}
\chi_\|({\bf k},\omega),\\
\label{dsfperp}
{\cal S}_\perp({\bf k},\omega) &=& 
\frac1\pi {\rm Im}
\chi_\perp({\bf k},\omega),
\end{eqnarray}
where $z$ axis is directed along staggered magnetizations,
\begin{equation}
\label{skstr}
	{\bf S}_{\bf k} = \frac{1}{2} 
	\left( 
	{\bf S}_{1\bf k} + {\bf S}_{2\bf k}e^{-ik_2/2} 
	+ {\bf S}_{3\bf k}e^{-i(k_1+k_2)/2} + {\bf S}_{4\bf k}e^{-ik_2/2}
	\right)
\end{equation}
are built on spin operators 1--4 in the extended unit cell (see Fig.~\ref{lattfig}(a)), ${\bf k} = k_1{\bf g}_1 + k_2{\bf g}_2$, and ${\bf g}_{1,2}$ are depicted in Fig.~\ref{lattfig}(b). 

\begin{figure}
\includegraphics[scale=0.9]{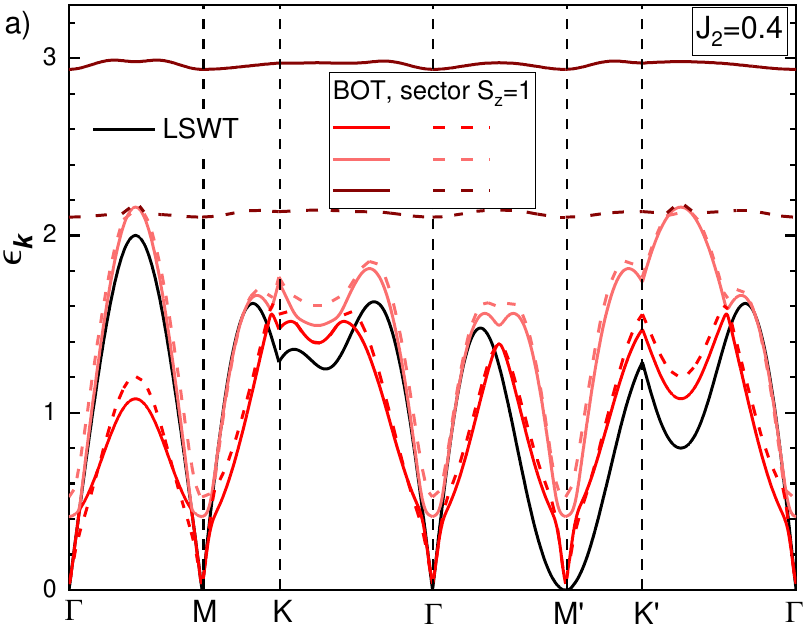}
\includegraphics[scale=0.9]{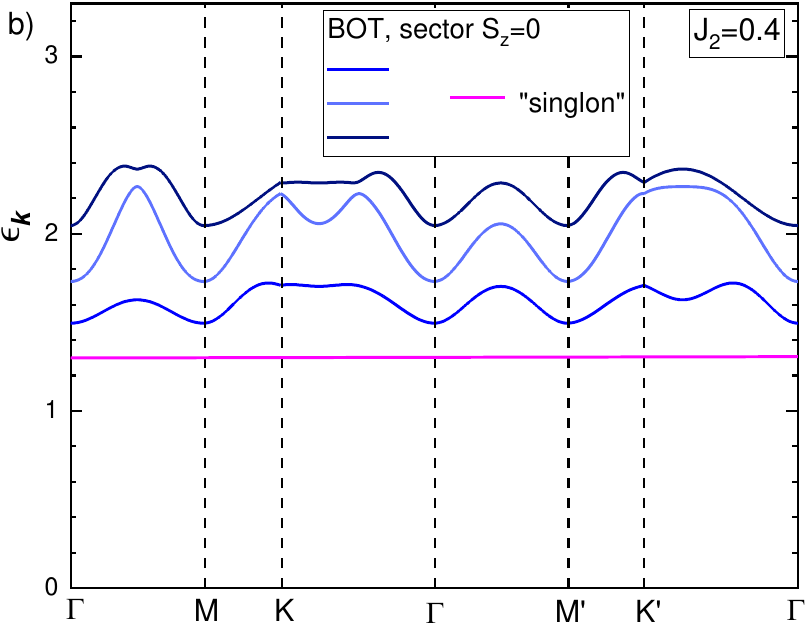}
\includegraphics[scale=0.9]{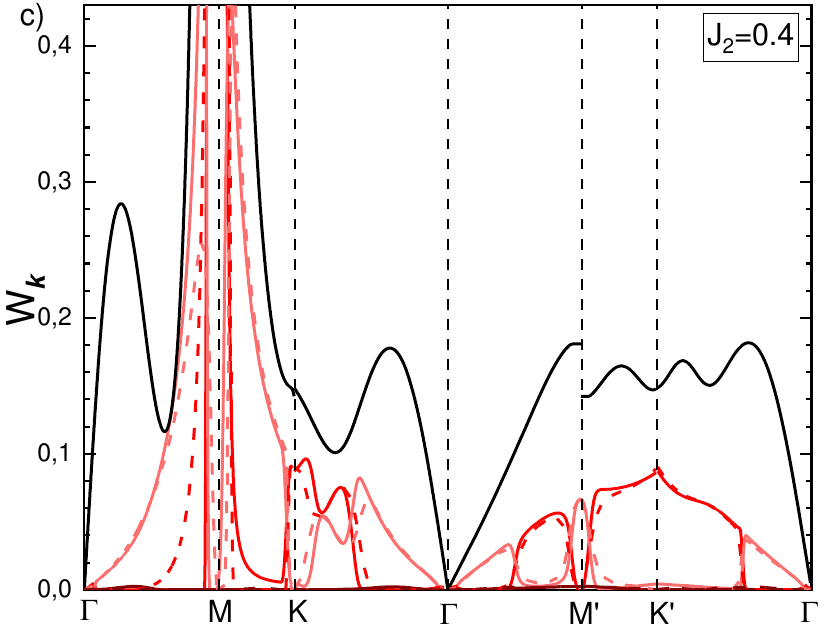}
\includegraphics[scale=0.9]{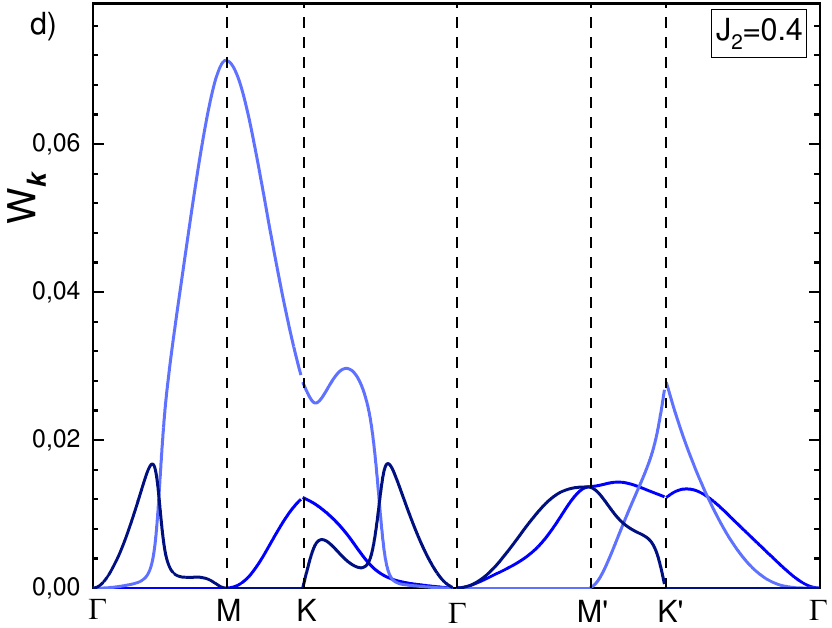}
\caption{
Spectra of elementary excitations arising in (a) transverse channel (the sector with the projection of the total spin $S_z=1$) and (b) longitudinal channel (the sector with $S_z=0$) which are obtained in the stripe phase at $J_2=0.4$ in the linear spin-wave theory (LSWT) and within the harmonic approximation of the BOT. All spin-0 and spin-1 quasiparticles correspond to poles of dynamical spin susceptibilities \eqref{chizz} and \eqref{chiperp}, respectively, with the exception of the spin-0 elementary excitation named "singlon" in panel (b) which corresponds to a pole of four-spin correlator \eqref{chia}. Spectral weights $W_{\bf k}$ of all quasiparticles are presented in panels (c) and (d) by colors corresponding to panels (a) and (b). $W_{\bf k}$ are coefficients before corresponding delta-functions in Eqs.~\eqref{dsfzz} and \eqref{dsfperp} (see Eq.~\eqref{sha}).
\label{spec004}}
\end{figure}

In the leading order in $1/n$, susceptibilities \eqref{chizz} and \eqref{chiperp} have the form of linear combinations of Green's functions of bosons. \cite{ibot} Then, DSFs \eqref{dsfzz} and \eqref{dsfperp} have the structure in the harmonic approximation of the BOT
\begin{equation}
\label{sha}
	{\cal S}^{(ha)}({\bf k},\omega) = \sum_i W_{i\bf k} \delta\left(\omega-\epsilon^{(0)}_{i\bf k}\right),
\end{equation}
where $i$ enumerates spectra branches, $\epsilon^{(0)}_{i\bf k}$ are bare quasiparticles spectra, and $W_{i\bf k}$ are their spectral weights. Elementary excitations corresponding to poles of spin susceptibilities \eqref{chizz} and \eqref{chiperp} carry spin 0 and 1, respectively. Their spectra found in the harmonic approximation of the BOT are shown in Figs.~\ref{spec004}(a) and \ref{spec004}(b), correspondingly, and their spectral weights are presented in Figs.~\ref{spec004}(c) and \ref{spec004}(d). It is interesting to compare these results with findings of the linear SWT which are also presented in Figs.~\ref{spec004}(a) and \ref{spec004}(c).  

It is well known that there is a doubly degenerate magnon spectrum in the two-sublattice stripe phase within the SWT which is zero at $\Gamma$ and $M$ points (notice that the $\Gamma M$ line is perpendicular to ferromagnetic chains, see Fig.~\ref{lattfig}). \cite{gsj1j21,gsj1j22} The classical spectrum is also zero at $M'$ point due to an accidental degeneracy of the ground state. \cite{gsj1j22} However, first $1/S$ corrections produce a gap at $M'$ point via the order-by-disorder mechanism. \cite{gsj1j22}

As is seen from Fig.~\ref{spec004}(a), there are six low-energy spin-1 excitations in the BOT two of which have zero energy at $\Gamma$ and $M$ points. However it is seen from Fig.~\ref{spec004}(c) that in the most part of the BZ only two of these spin-1 modes have predominant spectral weights in $\chi_\perp({\bf k},\omega)$ whose spectra follow the doubly degenerate magnon spectrum in the linear SWT (except for the close neighborhood of $M'$ point, where a portion of quantum fluctuations taken into account in the harmonic approximation of the BOT leads to gaps in two spin-1 modes having finite spectral weight at $M'$). It should be stressed that these two spin-1 modes are split in the BOT that is the effect of quantum fluctuations more accurately taken into account in our approach than in the SWT, where the magnon spectrum remains degenerate even in the first order in $1/S$ \cite{gsj1j21,gsj1j22}. Similar magnon modes splitting by quantum fluctuations is observed above in the $120^\circ$ phase. Notice that the BOT does not tend to split all degenerate branches: spin-1 modes remain doubly degenerate in the similar two-sublattice collinear phase of the Heisenberg antiferromagnet on the square lattice considered by the four-spin version of the BOT in our previous papers \cite{ibot,aktersky}. It is shown below that this splitting vanishes at $J_2=1$, where the transition to the spiral phase takes place, and that first corrections in $1/n$ enhance the difference in the stripe state between branches shown in Fig.~\ref{spec004}(a) in dashed and solid lines of the same color.

It is also seen from Figs.~\ref{spec004}(a) and \ref{spec004}(c) that the translation symmetry is restored in the four-spin version of the BOT which uses the artificially enlarged unit cell: despite energies of each quasiparticle are equal at equivalent points of the reciprocal space built on vectors ${\bf g}_{1,2}$ (see Fig.~\ref{lattfig}(b)), spectral weights differ at such points if they are not equivalent in the scheme with two spins in the unit cell (in the latter case the reciprocal space is built on vectors ${\bf b}_2$ and  $({\bf b}_1+{\bf b}_2)/2$). For instance, one concludes from Figs.~\ref{spec004}(c) that two pink and two red branches have predominant spectral weights at points $Y$ and $X'$, respectively (although $Y$ and $X'$ are equivalent in the scheme with the four-site unit cell).

Branches of two high-energy spin-1 excitations shown in brown in Fig.~\ref{spec004}(a) can arise in the SWT as bound states of three magnons. Although their spectral weights are small in the harmonic approximation of the BOT (see Fig.~\ref{spec004}(c)), they produce visible anomalies in DSFs in the first order in $1/n$ as it is demonstrated below.

\begin{figure}
\includegraphics[scale=0.99]{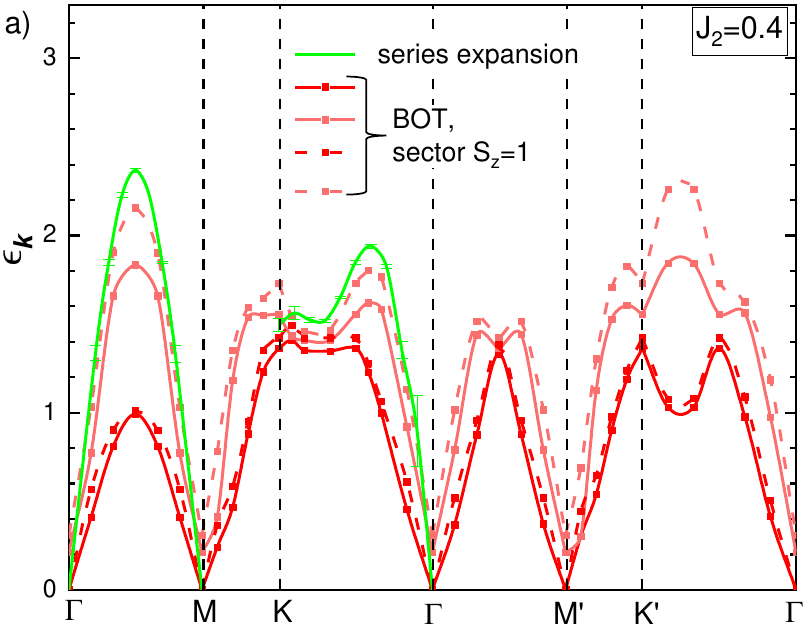}
\includegraphics[scale=0.99]{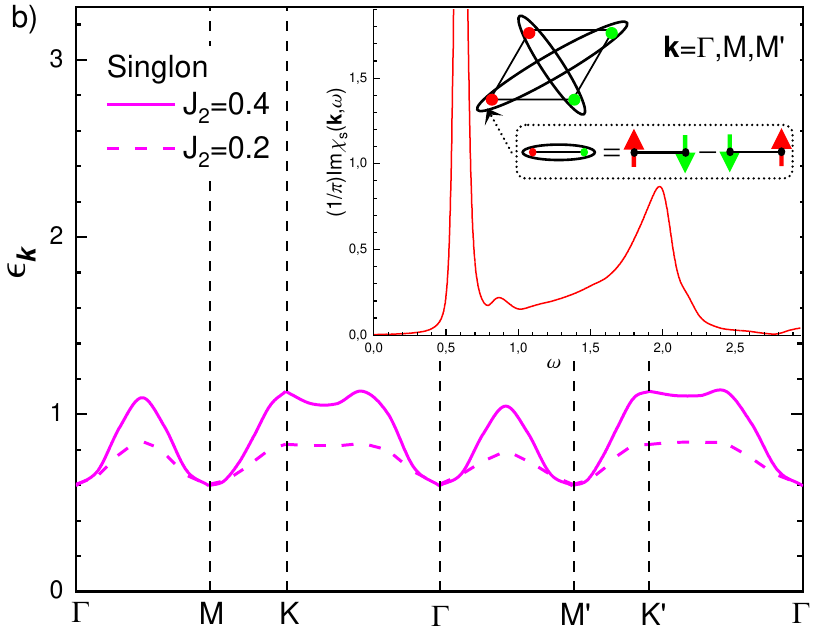}
\caption{(a) Spectra of low-energy spin-1 quasiparticles found in the first order in $1/n$ (BOT). Corresponding spectra in the harmonic approximation are shown in Fig.~\ref{spec004}(a). Two high-energy branches presented in Fig.~\ref{spec004}(a) acquire very large damping in the first order in $1/n$ and they are not shown here. Series expansion data are taken from Ref.~\cite{oitmaatri}. (b) Spectrum of spin-0 excitation named "singlon" and discussed in the text which was found self-consistently in the first order in $1/n$ at $J_2=0.2$ and 0.4. Inset shows DSF $\frac1\pi {\rm Im}\chi_s({\bf k},\omega)$ at $J_2=0.4$ built on four-spin susceptibility \eqref{chia} at $\Gamma$, $M$, and $M'$ points and convoluted with the energy resolution of $0.03$. Another inset presents the singlet state of the four-spin unit cell created by the Bose operator describing the singlon in the harmonic approximation of the BOT.
\label{specren04}}
\end{figure}

Spectra of spin-1 excitations found at $J_2=0.4$ in the first order in $1/n$ are shown in Fig.~\ref{specren04}(a). 
It is seen that spectra of two branches having the largest spectral weights are close to magnon energies found in Ref.~\cite{oitmaatri} by the series expansion. Two high-energy branches presented in Fig.~\ref{spec004}(a) acquire very large damping in the first order in $1/n$ and they are not presented in Fig.~\ref{specren04}(a). However, at some part of the BZ, these two quasiparticles have well-defined spectra found in the self-consistent scheme discussed above. 

\begin{figure}
\includegraphics[scale=0.73]{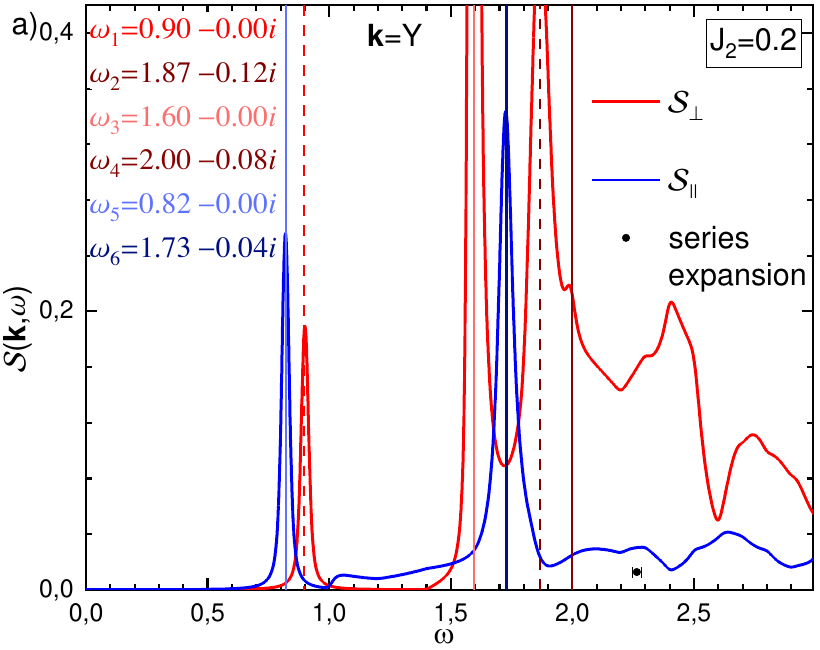}
\includegraphics[scale=0.73]{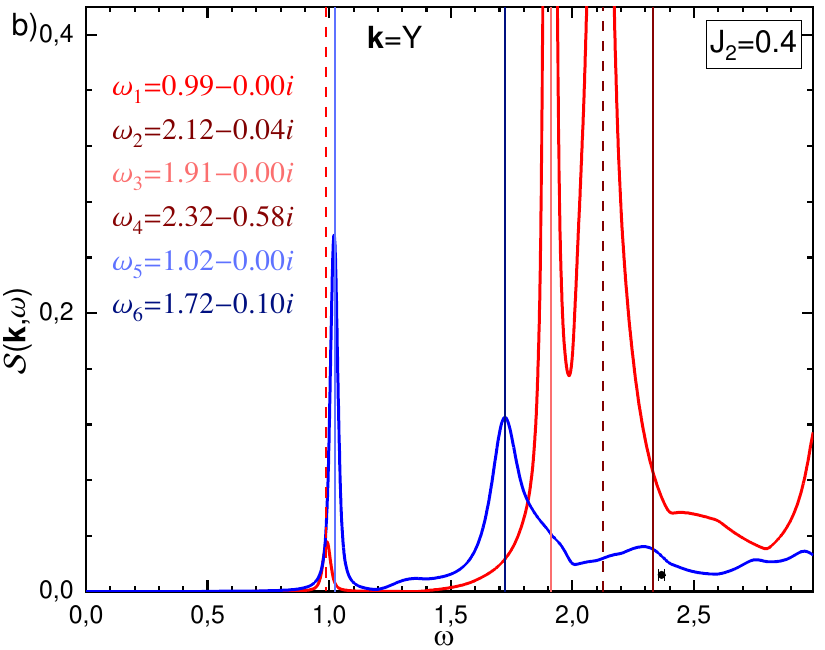}
\includegraphics[scale=0.73]{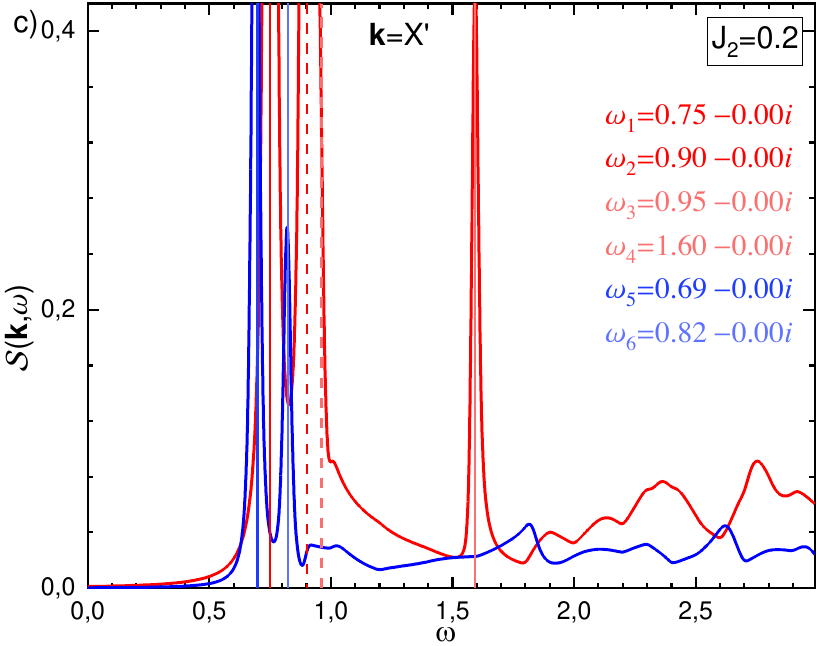}
\includegraphics[scale=0.73]{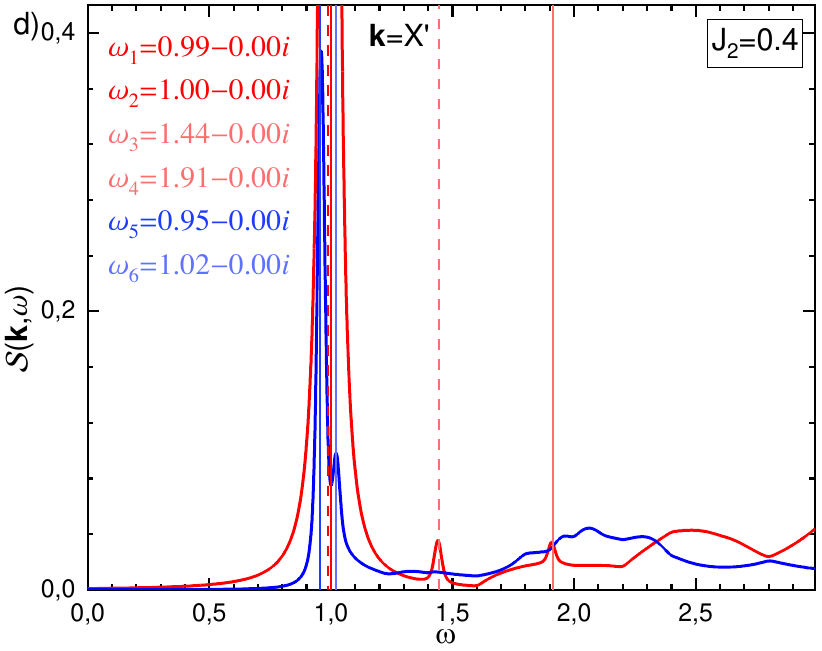}
\includegraphics[scale=0.73]{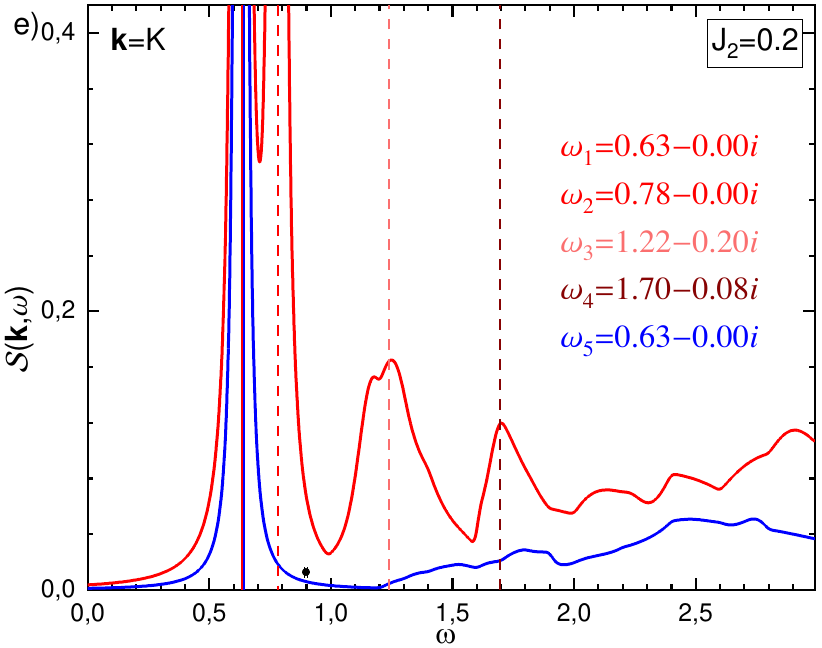}
\includegraphics[scale=0.73]{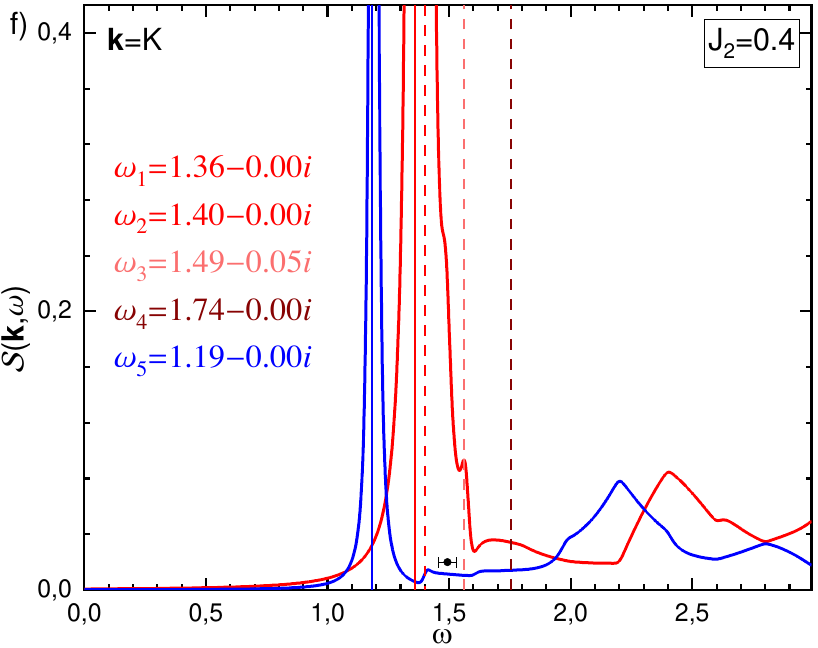}
\includegraphics[scale=0.73]{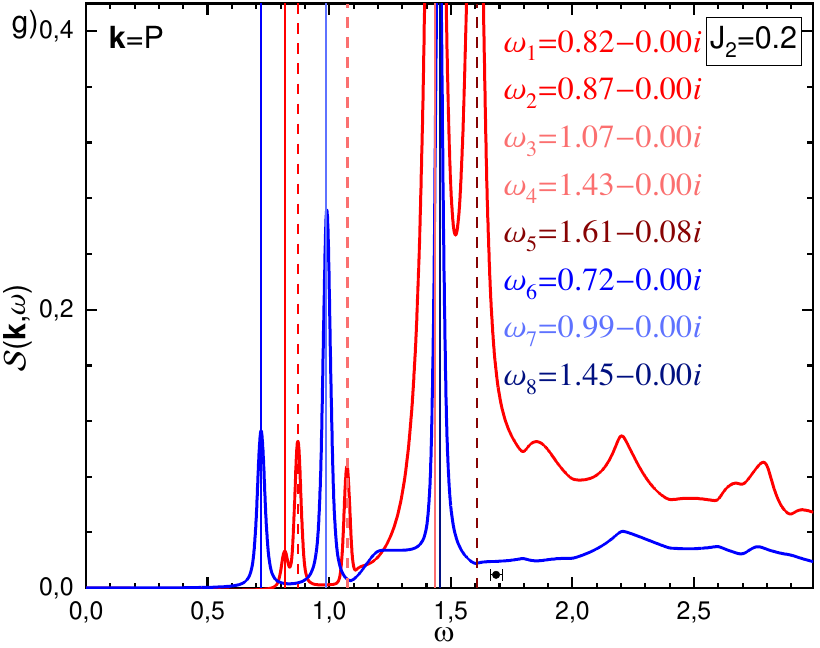}
\includegraphics[scale=0.73]{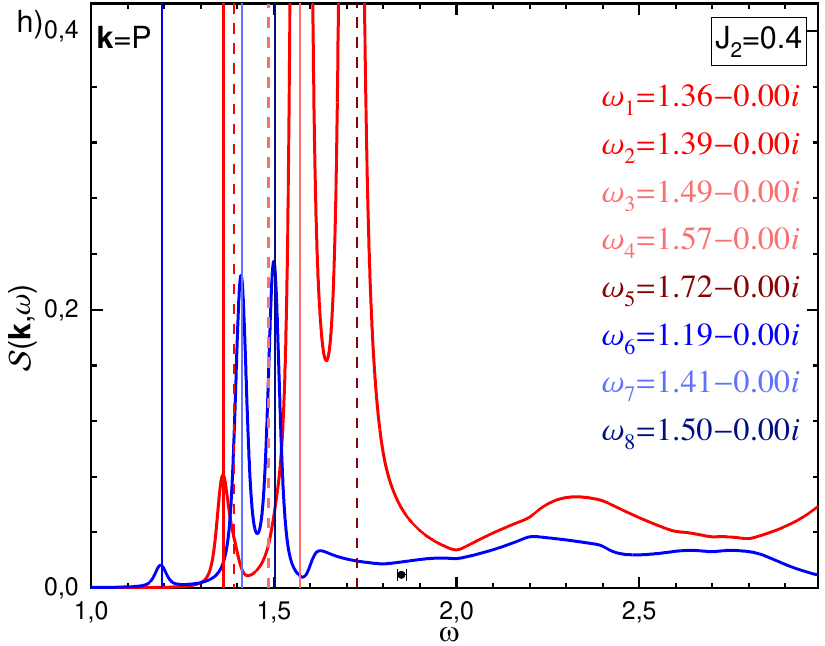}
\includegraphics[scale=0.73]{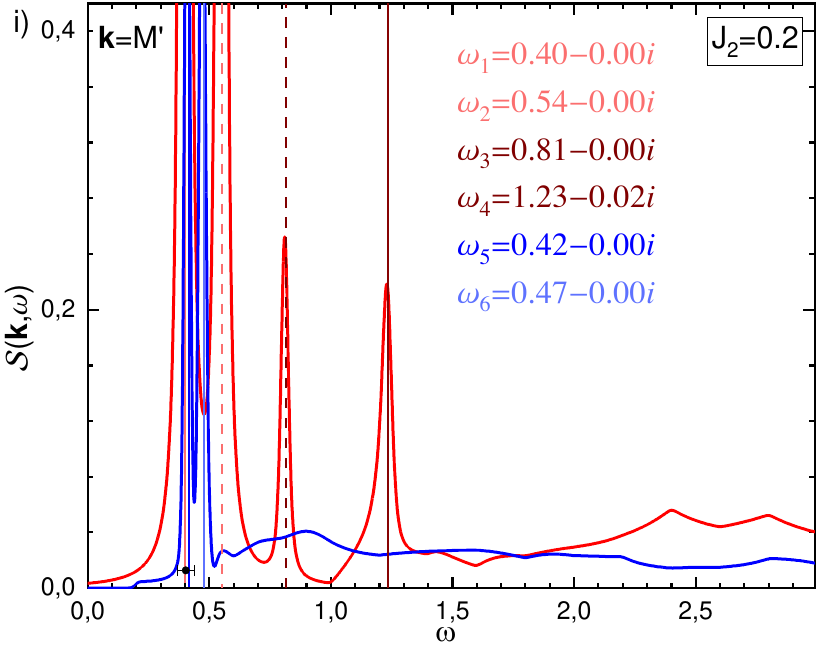}
\includegraphics[scale=0.73]{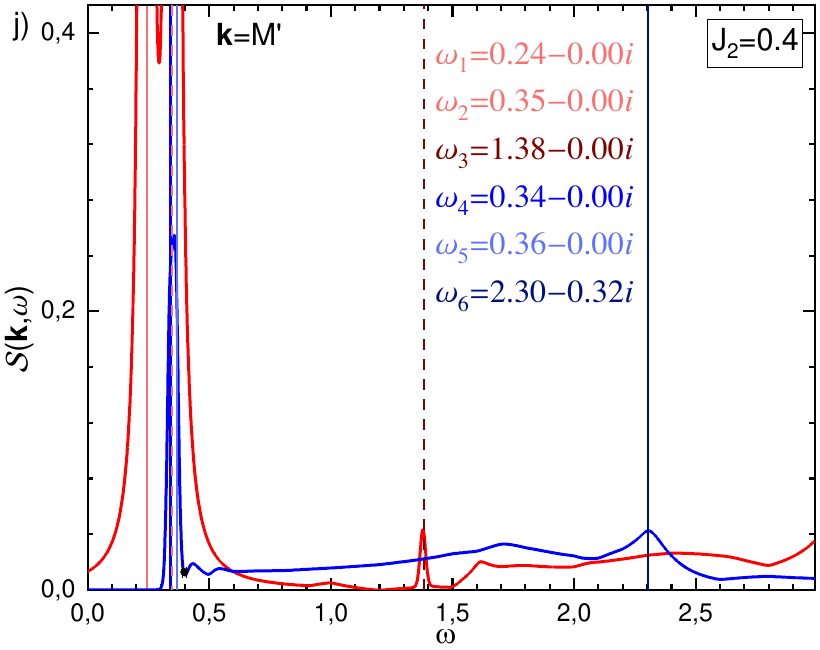}
\caption{
Same as Fig.~\ref{dsfmy} but for the transverse ${\cal S}_\perp({\bf k},\omega)$ \eqref{dsfperp} and the longitudinal ${\cal S}_\|({\bf k},\omega)$ \eqref{dsfzz} DSFs at points $Y$, $X'$, $K$, $P$, and $M'$ of the BZ (see Fig.~\ref{lattfig}(b)) in the stripe phase at $J_2=0.2$ and $J_2=0.4$. DSFs obtained within the first order in $1/n$ have been convoluted with the energy resolution of $0.01$.
\label{dsfstr04att}}
\end{figure}

Fig.~\ref{dsfstr04att} illustrates this our finding, where DSFs are presented obtained in the first order in $1/n$ at points $Y$, $X'$, $K$, $P$, and $M'$ and poles are shown in insets which correspond to considered quasiparticles and which are found self-consistently. Notice that energies of four low-energy spin-1 excitations found self-consistently differ little from spectra in the first order in $1/n$ (cf.\ Fig.~\ref{specren04}(a) and insets in Figs.~\ref{dsfstr04att}(b), \ref{dsfstr04att}(d), \ref{dsfstr04att}(f), \ref{dsfstr04att}(h), and \ref{dsfstr04att}(j)). Interestingly, two high-energy spin-1 elementary excitations (shown in brown) produce pronounced anomalies at $Y$ and $P$. The large difference between their spectra found in the first order in $1/n$ and self-consistently may also indicate the need to go beyond the first order in $1/n$ to find their spectra accurately. 

Notice also that $1/n$ corrections increase the splitting between branches shown in Fig.~\ref{spec004}(a) by the same color (see Fig.~\ref{specren04}(a) and insets in Fig.~\ref{dsfstr04att}). Moreover, the renormalization of spectral weights of the split bands by $1/n$ corrections is very different so that quasiparticles from branches of the same color can appear separately in DSFs (see Fig.~\ref{dsfstr04att}). It is also seen from Fig.~\ref{dsfstr04att} that quasiparticles from branches of different colors which appear simultaneously in DSFs in the first order in $1/n$ have substantially different spectral weights. That is why we cannot state that all four low-energy excitations can appear simultaneously at some points of the BZ. We point out also a very different physical picture at $Y$ and $X'$ points (see Figs.~\ref{dsfstr04att}(a), \ref{dsfstr04att}(b), \ref{dsfstr04att}(c), and \ref{dsfstr04att}(d)) which are equivalent in the scheme with the four-spin unit cell.

There are three low-energy spin-0 excitations in the longitudinal channel whose spectra obtained in the harmonic approximation of the BOT are shown in Fig.~\ref{spec004}(b). To the best of our knowledge, these quasiparticles have not been discussed yet in the stripe phase by other approaches. It is seen from Fig.~\ref{spec004} that their spectral weights are quite comparable with those of spin-1 quasiparticles except for the vicinity of $M$ point and their energies are close to high-energy parts of magnon spectra. Renormalization of their spectra by first $1/n$ corrections are large as in the case of two high-energy spin-1 excitations discussed above. Then, we present here only the longitudinal DSF for five points in the BZ at $J_2=0.2$ and $J_2=0.4$ (see Fig.~\ref{dsfstr04att}). It is seen from Fig.~\ref{dsfstr04att} that spectra of almost all spin-0 excitations found self-consistently have very small damping. Besides, spectral weights of anomalies in the longitudinal DSF originating from spin-0 quasiparticles rise upon approaching the SLP near which they are comparable with spectral weights of peaks produced by spin-1 quasiparticles in the transverse DSF. Energies of low-energy spin-0 excitations are close to energies of low-energy spin-1 quasiparticles near the QCP. Possibly, some spin-0 and spin-1 quasiparticles introduced here merge at the QCP forming triplon excitations.

There is a special spin-0 quasiparticle corresponding to a pole not in $\chi_\|({\bf k},\omega)$ but in the four-spin correlator
\begin{equation}
\label{chia}
\begin{aligned}
&\chi_s({\bf k},\omega) =
i\int_0^\infty dt 
e^{i\omega t}	
\left\langle \left[ {\cal A}_{\bf k}(t), {\cal A}^\dagger_{-\bf k}(0) \right] \right\rangle,\\
&{\cal A}_j = {\bf S}_{1j}{\bf S}_{3j} - {\bf S}_{2j}{\bf S}_{4j},
\end{aligned}
\end{equation}
where ${\bf S}_{pj}$ is the $p$-th spin in the $j$-th unit cell. In our previous consideration of the similar two-sublattice collinear phase in the Heisenberg antiferromagnet on the square lattice \cite{ibot}, we call this elementary excitation "singlon" because Bose operator describing it creates a singlet spin state of the four-spin unit cell in the harmonic approximation of the BOT (see inset in Fig.~\ref{specren04}(b)). We show in Ref.~\cite{ibot} that singlon produces a broad peak in the Raman intensity in the $B_{1g}$ geometry which was observed, in particular, in layered cuprates. The spectrum of singlon is dispersionless in the harmonic approximation (see Fig.~\ref{spec004}(b)) but first $1/n$ corrections lead to the dispersion as it is seen from Fig.~\ref{specren04}(b). Notice that there is no damping in the singlon spectrum found self-consistently and shown in Fig.~\ref{specren04}(b). Interestingly, the singlon spectrum lies below energies of all spin-0 and spin-1 excitations in some parts of the BZ. The DSF built on four-spin susceptibility \eqref{chia} is shown in the inset of Fig.~\ref{specren04}(b) at $\Gamma$, $M$, and $M'$ points.

It is known that $\rm KCeS_2$ is described by model \eqref{ham} in the stripe state. However an experimental test of our predictions is impossible because neutron data obtained only on powder samples are available now. \cite{kces}

\subsection{Transition to the spiral phase}
\label{spiral}

\begin{figure}
\includegraphics[scale=0.9]{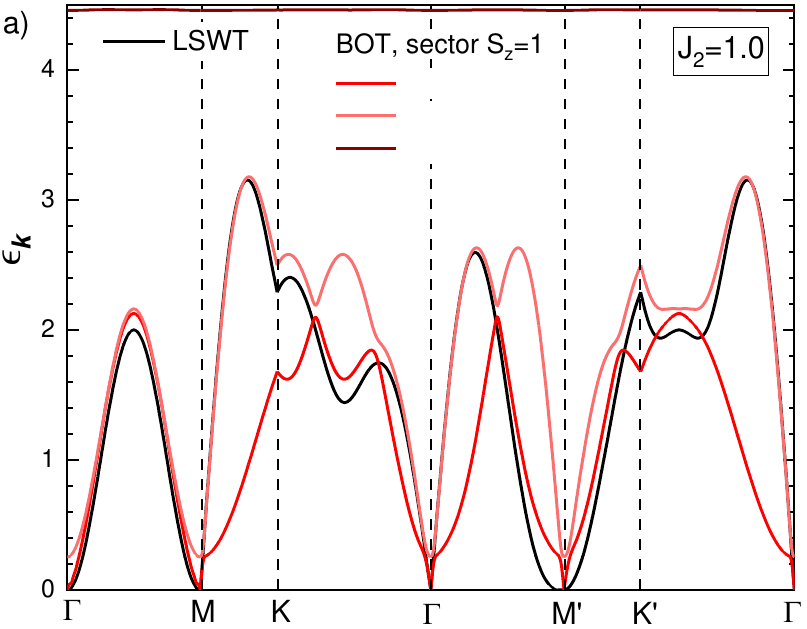}
\includegraphics[scale=0.9]{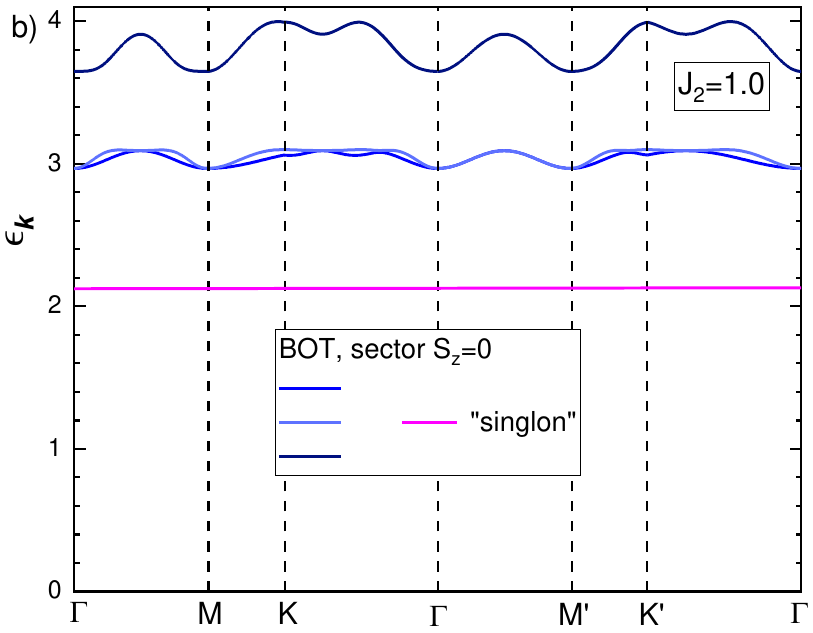}
\includegraphics[scale=0.9]{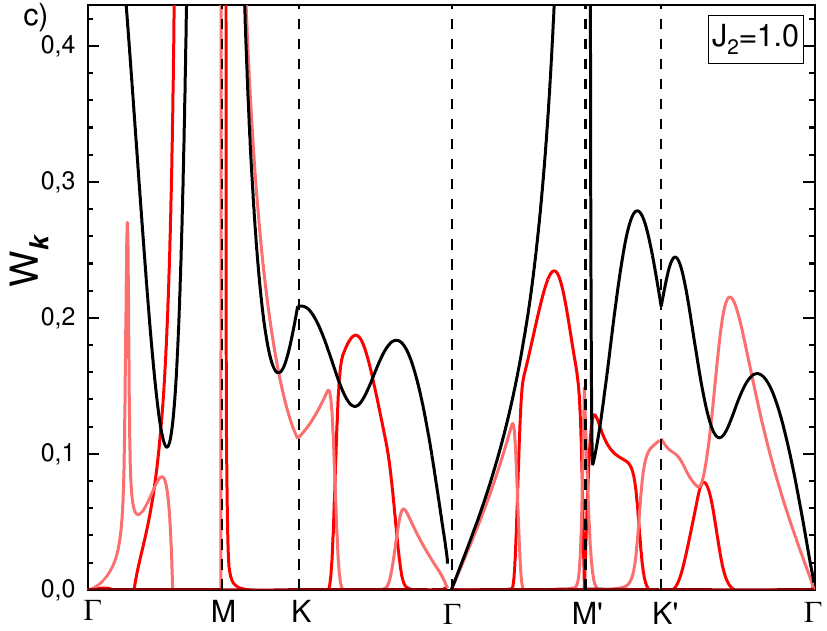}
\includegraphics[scale=0.9]{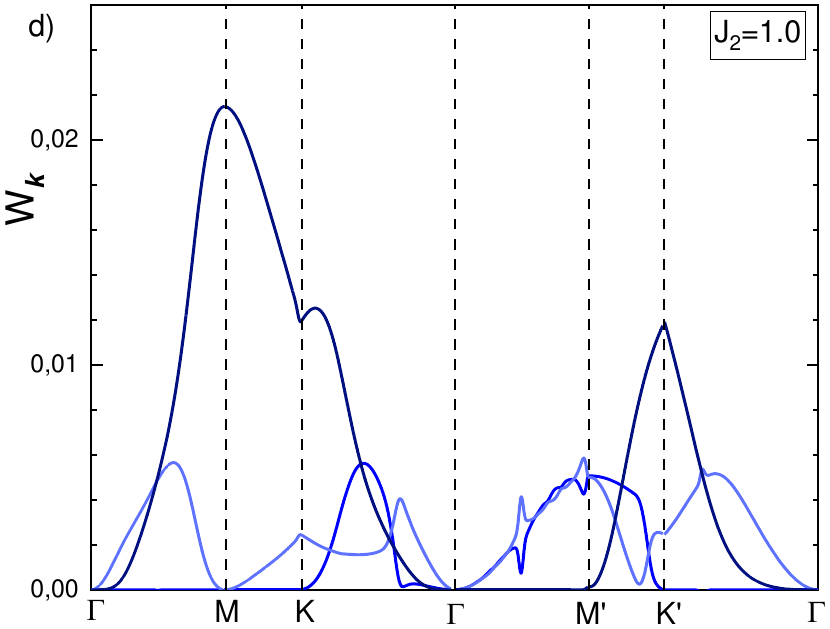}
\caption{
Same as Fig.~\ref{spec004} but for $J_2=1$.
\label{spec010}}
\end{figure}

Within the harmonic approximation of the BOT, the transition to the spiral order occurs at $J_2=1$ as in the LSWT (see Ref.~\cite{gsj1j22}). Within both approaches, the velocity of the Goldstone magnon vanishes in the direction perpendicular to the ferromagnetic chains at $J_2=1$. As it is seen from Fig.~\ref{spec010}(a), there are three doubly degenerate spin-1 modes in the BOT at $J_2=1$ which are split at $J_2<1$ (cf.\ Figs.~\ref{spec004}(a) and \ref{spec010}(a)). Fig.~\ref{spec010}(b) shows that energies of all spin-0 quasiparticles move up upon approaching the transition to the spiral phase.

Further consideration of the spiral phase with the incommensurate magnetic order is not simple within the BOT and it is out of the scope of the present paper.

\section{Summary and Conclusion}
\label{conc}

To conclude, we discuss dynamics of spin-$\frac12$ $J_1$--$J_2$ model \eqref{ham} on the triangular lattice using the bond-operator theory (BOT) proposed in Refs.~\cite{ibot,itri}. We calculate spectra of elementary excitations and dynamical structure factors (DSFs) in the first order in $1/n$ in the $120^\circ$ and in the stripe ordered states (see Fig.~\ref{lattfig}(a)). All calculated static characteristics of the model are in good agreement with previous numerical findings (see Fig.~\ref{ME}). Domain of stability of the spin-liquid phase (SLP) \eqref{slreg} is also in a good agreement with previous numerical results.

In the $120^\circ$ phase, we observe the evolution of quasiparticles spectra and DSF \eqref{dsf} upon approaching the SLP (see Fig.~\ref{dsfmy}). We demonstrate strong modification by quantum fluctuations of conventional magnons which is not captured by the semi-classical spin-wave theory (SWT). Other considered elementary excitations were introduced first in our previous paper \cite{itri} devoted to model \eqref{ham} at $J_2=0$. All obtained quasiparticles produce visible anomalies in the DSF. We demonstrate that the continuum of excitations moves closer to the lowest well-defined magnon mode upon $J_2$ increasing in agreement with previous numerical findings \cite{triang3}. The remaining two conventional magnon modes acquire noticeable damping on the way to the SLP while some other high-energy modes found in Ref.~\cite{itri} remain well-defined and produce visible anomalies in the DSF. Our results are in overall agreement with neutron data obtained in $\rm KYbSe_2$.

In the stripe phase, we observe that the doubly degenerate magnon spectrum known from the SWT is split by quantum fluctuations which are taken into account more accurately in the BOT. This splitting vanishes at the transition point to the spiral phase (at $J_2=1$). Similar splitting of two magnon branches is observed by the BOT in the $120^\circ$ phase that is in quantitative agreement with experimental data obtained in $\rm Ba_3CoSb_2O_9$. \cite{itri} As compared with other known results of the SWT, we observe additional spin-0 and spin-1 quasiparticles which give visible anomalies in the longitudinal and transverse DSFs (see Fig.~\ref{dsfstr04att}) and which would appear in the SWT as bound states of two and three magnons, respectively. Energies of lowest spin-0 quasiparticles become closer to energies of lower spin-1 excitations upon approaching the SLP. Spectral weights of peaks produced by well-defined spin-0 excitations are also increased on the way to the SLP (see Fig.~\ref{dsfstr04att}). We observe also a special well-defined spin-0 quasiparticle named singlon who produces a peak only in four-spin correlator \eqref{chia}. Singlon is invisible in the longitudinal DSF but its spectrum at zero momentum can be probed by the Raman scattering. We find that the singlon spectrum lies below spectra of all spin-0 and spin-1 excitations in some parts of the Brillouin zone (see Fig.~\ref{specren04}(b)).

We hope that our results will stimulate further theoretical and experimental activity in this field. 

\begin{acknowledgments}

I am grateful to A.~Trumper for useful discussion and the exchange of data. This work is supported by the Russian Science Foundation (Grant No.\ 22-22-00028). 

\end{acknowledgments}

\bibliography{tribib}

\end{document}